\documentclass[aps,pre,twocolumn,amsmath,amsfonts,floatfix,superscriptaddress]{revtex4}
\usepackage{graphicx} 

 \let\b=\beta \let\g=\gamma \let\d=\delta
  \let\h=\eta 
 \let\m=\mu \let\n=\nu  \let\p=\pi
  \let\f=\varphi 
   \let\G=\Gamma
   
\let\Si=\Sigma   
 \let\r=\rho  

\def\ie{{i.e. }}

\def\NN{{\cal N}} 
  
 \def\SS{{\cal S}}

\def\to{\rightarrow} \def\la{\left\langle} \def\ra{\right\rangle}

\newcommand{\beq}{\begin{equation}} \newcommand{\eeq}{\end{equation}}
 \newcommand{\wt}{\widetilde}

\def\piero{paper }
\def\aldo{Appendix}

\begin{document}

\title{
A theory of amorphous packings of binary mixtures of hard spheres
}
\author{Indaco Biazzo}

\affiliation{Dipartimento di Fisica, Universit\`a di Roma ``La Sapienza'', 
P.le A. Moro 2, 00185 Roma, Italy}

\author{Francesco Caltagirone}

\affiliation{Dipartimento di Fisica, Universit\`a di Roma ``La Sapienza'', 
P.le A. Moro 2, 00185 Roma, Italy}

\author{Giorgio Parisi}

\affiliation{Dipartimento di Fisica, Universit\`a di Roma ``La Sapienza'', 
P.le A. Moro 2, 00185 Roma, Italy}

\affiliation{INFM-CNR SMC, INFN,
Universit\`a di Roma ``La Sapienza'',
P.le A. Moro 2, 00185 Roma, Italy
}

\author{Francesco Zamponi}

\affiliation{Laboratoire de Physique Th\'eorique, CNRS UMR 8549,
Ecole Normale Sup\'erieure,
24 Rue Lhomond, 75231 Paris Cedex 05, France
}


\begin{abstract} 
We extend our theory of amorphous packings of hard spheres 
to binary mixtures and more generally to multicomponent systems.
The theory is based on the assumption that amorphous packings
produced by typical experimental or numerical protocols
can be identified with the infinite pressure limit of long lived
metastable glassy states.
We test this assumption against numerical and experimental data
and show that the theory correctly reproduces
the variation with mixture composition of structural observables, such as the total
packing fraction and the partial coordination numbers.
\end{abstract} 


\maketitle

Amorphous packings of hard spheres are ubiquitous in physics: 
they have been used as models for liquids, glasses,
colloidal systems, granular systems, and powders. 
They are also related to important problems in mathematics and 
information theory, such as digitalization of
signals, error correcting codes, and optimization problems. 
Moreover, the structure and density (or porosity) of 
amorphous multicomponent packings is
important in many branches of science and technology,
ranging from oil extraction to storage of grains in silos.

Despite being empirically studied since at least sixty years,
amorphous packings still lack a precise mathematical definition, due
to the intrinsic difficulty of quantifying ``randomness''~\cite{TTD00}.
Indeed, even if a sphere packing is a purely geometrical object,
in practice dense amorphous packings always result from
rather complicated dynamical protocols: for instance, spheres can be
thrown at random in a box that is subsequently shaken to achieve compactification~\cite{SK69},
or they can be deposited onto a random seed 
cluster~\cite{Be72}.
In numerical simulations,
one starts from a random distribution of small spheres and inflates them
until a jammed state is reached~\cite{LS90,DTS06}; 
alternatively, one starts from large overlapping
spheres and reduces the diameter in order to eliminate the 
overlaps~\cite{JT85,CJ93,OLLN02,LAEMS06}.
In principle, each of these dynamical prescriptions produces an {\it ensemble}
of final packings that depends on the details of the procedure used.
Still, very remarkably, if the presence of crystalline regions is avoided, 
the structural properties of amorphous packings turn out to be very similar.
This observation led to the proposal that ``typical'' amorphous packings should
have common structure and density; the latter has been denoted
{\it Random Close Packing} (RCP) density.
The definition of RCP has been intensively debated in the
last few years, in connection with the progresses of numerical simulations~\cite{TTD00,KL07}.

Nevertheless, the empirical evidence, that amorphous packings produced according to very different
protocols have common structural properties, is striking and call for an explanation. 
This is all the more true for binary or multicomponent mixtures, where in addition to the
usual structural observables, such as the structure factor, one can investigate other
quantities such as the coordination between spheres of different type, and study
their variation with the composition of the mixture.

In earlier attempts to build statistical models of packings, 
only the main geometrical factors, such as the relative size
and abundance of the different components, were taken 
into account~\cite{Do75,Do80,OT81}. 
More precisely, these models focus on
a random sphere in the packing and its first neighbors, 
completely neglecting spatial correlations beside the first shell and all 
the global geometric constraints. 
This already accounts for the main
qualitative structural properties of random packings. However, in order to obtain
a quantitative description, some free parameters have to be introduced and adjusted
to match with experimental data.

To go beyond these simple models,
many authors proposed that random packings of hard spheres can be thought as
the infinite pressure limit of hard sphere glasses~\cite{WA81,SW84,Sp98,CFP98,PZ05,KK07,PZ08}.
This is very intuitive since a glass is a solid state in which particles 
vibrate around amorphous reference positions, and vibrations are reduced on increasing pressure.
A typical algorithm attempting to create a random packing starts
at low density and compresses the system at a given rate.
During this evolution, when the density is high enough, 
relaxation becomes more and more difficult until at some point the system is stuck into
a glass state~\cite{KK07,PZ08}; at this point further compression will only reduce the amplitude of the
vibrations. 
In a nutshell, this is why amorphous packings can be identified with glasses at infinite pressure.

The main advantage of this identification, if it holds, is that a glass is a {\it metastable state}
that has a {\it very} long lifetime; therefore, its properties 
can be studied using concepts of equilibrium statistical mechanics. In this way
a complicated {\it dynamical} problem (solving the equations of motion for a given protocol)
is reduced to a much more simple equilibrium problem.
In~\cite{PZ05,PZ08} it was shown, in the case of monodisperse packings, 
that this strategy is very effective since it allows
to compute structural properties of random packings directly from the Hamiltonian of the
system, without free parameters and in a controlled statistical mechanics framework.
Note that the existence of an {\it equilibrium} glass transition in hard sphere systems
(or in other words the existence of glasses with {\it infinite} life time)
has been questioned~\cite{DTS06,SK00}. Although very interesting, this problem is not relevant
for the present discussion since we are only interested in long-lived metastable glasses that trap dynamical
algorithms. At present it is very well established by numerical simulations~\cite{BW08,BW09} 
that for system sizes of $N\lesssim 10^4$ particles and on the time scales of typical
algorithms, metastable glassy states exist, at least in $d \geq 3$. This is enough to compare
with most of the currently available numerical and experimental data.
Finally, the relation of this approach to special packings such as the MRJ state~\cite{TTD00}
and the J-point~\cite{OLLN02} has been discussed in detail in \cite{PZ08}.

The aim of this \piero is to extend the theory of \cite{PZ05,PZ08} to binary mixtures.
This allows to compare quantitatively the predictions of the theory and the results of numerical
simulations. We will focus in particular on the variation of density and local connectivity
as a function of mixture composition. These results constitute, in our opinion, 
a stringent test of the assumption that random packings reached by standard algorithms 
can be identified with infinite pressure metastable glasses.

{\it Theory} -
The equilibrium statistical mechanics computation of the properties
of the glass is based on standard liquid theory~\cite{Hansen} and on the replica 
method~\cite{CFP98,MP99} that has
been developed in the context of spin glass theory~\cite{MPV87,Mo95}. 
For monodisperse hard spheres,
it has been described in great detail in \cite{PZ08}. The extension to multicomponent
systems is straightforward following \cite{CMPV99}; details are given
in \aldo.

Here we just recall some features of this approach,
based on simple physical considerations.
The basic assumption of the method are that
{\it i)} crystallization and phase separation are strongly suppressed by kinetic
effects so that the liquid can be safely followed at high density, and {\it ii)}
that at sufficiently high density, 
the liquid is a superposition of a collection of amorphous {\it metastable}
states. Namely, in the liquid, the system spends some time inside one
of these states, and sometimes undergoes a rearrangement 
that leads to a different state~\cite{Go69}.
Each state is characterized by its vibrational entropy per particle, denoted by $s$,
and the number of such states is assumed to be exponential in $N$, so defining a
{\it configurational entropy} $\Si(\f,s) = N^{-1} \log \NN(\f,s)$, being $\NN(\f,s)$ the
number of states having entropy $s$ at density $\f$.
On increasing the density, the liquid is trapped for longer and longer times into
a metastable state, until at some point the transition time becomes so long that for all
practical purposes the system is stuck into one state: it then becomes a glass.
To compute the properties of the glassy states, the central problem is
to compute the function
$\Si(\f,s)$. This can be done by means of a simple replica method introduced by
Monasson \cite{Mo95}. One introduces $m$ copies of each
particle, constrained to be close enough, 
in such a way that they must be in
the same metastable state.
Then, the total entropy of the system of $m$ copies is given
by $\SS(m,\f | s) = m s + \Si(\f,s)$; the first term gives the entropy
of the $m$ copies in a state of entropy $s$, while the second term is due to the 
multiplicity of possible states.
The total entropy of the system at fixed density $\f$ is obtained by
maximizing over $s$, \ie
\beq
\SS(m,\f) = \max_s [\Si(\f,s) + m s] = \Si(\f,s^*) + m s^* \ ,
\eeq
where $s^*(m,\f)$ is determined by the condition $\partial_s \Si(\f,s) = m$.
Then it immediate to show that
\beq\label{legm}
\begin{split}
&s^*(m,\f) = \partial_m \SS(m,\f) \ , \\
&\Si(\f,s^*(m,\f)) = - m^2 \partial_m [\SS(m,\f)/m] \ .
\end{split}\eeq
The knowledge of $\SS(m,\f)$ allows to reconstruct the curve
$\Si(\f,s)$ for a given density by a parametric plot of
Eqs.~(\ref{legm}) by varying $m$. 
The function $\Si(\f,s)$ gives access to the internal entropy and the number  
of metastable glassy states; from this one can compute their equation of state,
\ie the pressure as a function of the density; in particular, for each set of
glassy states of given configurational entropy $\Si_j$, one can compute the density
$\f_j$ ({\it jamming density}) at which their pressure diverges.
Since $\f_j$ turns out to depend (slightly) on $\Si_j$, a prediction of the theory
is that {\it different glasses will jam at different density}: amorphous packings
can be found in a finite (but small) interval of density~\cite{KK07,PZ08}.

\begin{figure}
\includegraphics[width=8cm]{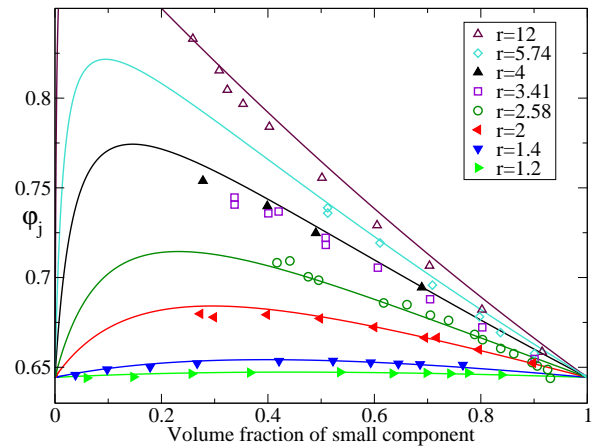}
\caption{
(Color online)
Packing fraction $\f_j$ as a function of 
$\eta = 1/(1+x r^3)$ at fixed $r$.
Full symbols are numerical data from this work. Open symbols are experimental results from Ref.~\cite{YCW65}. 
Lines are predictions from theory,
obtained fixing $\Si_j=1.7$. Note that the large $r$-small $\h$ region could not be explored, since for
such very asymmetric mixtures the large spheres form a rigid structure while
small spheres are able to move through the pores and are not jammed~\cite{Do80,OT81}.
}
\label{fig:phij}
\end{figure}

\begin{figure*}
\includegraphics[width=7cm]{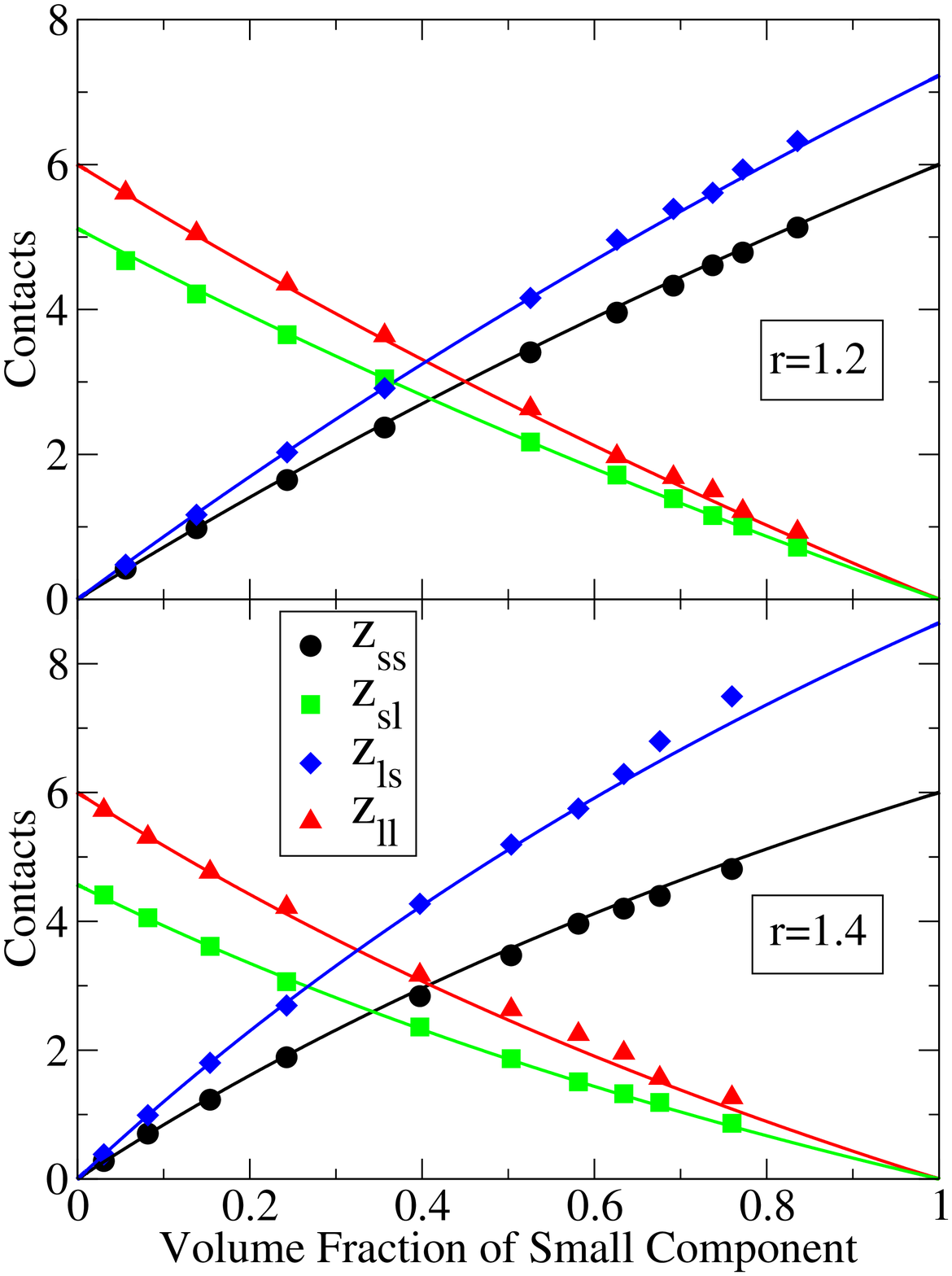}
\includegraphics[width=7cm]{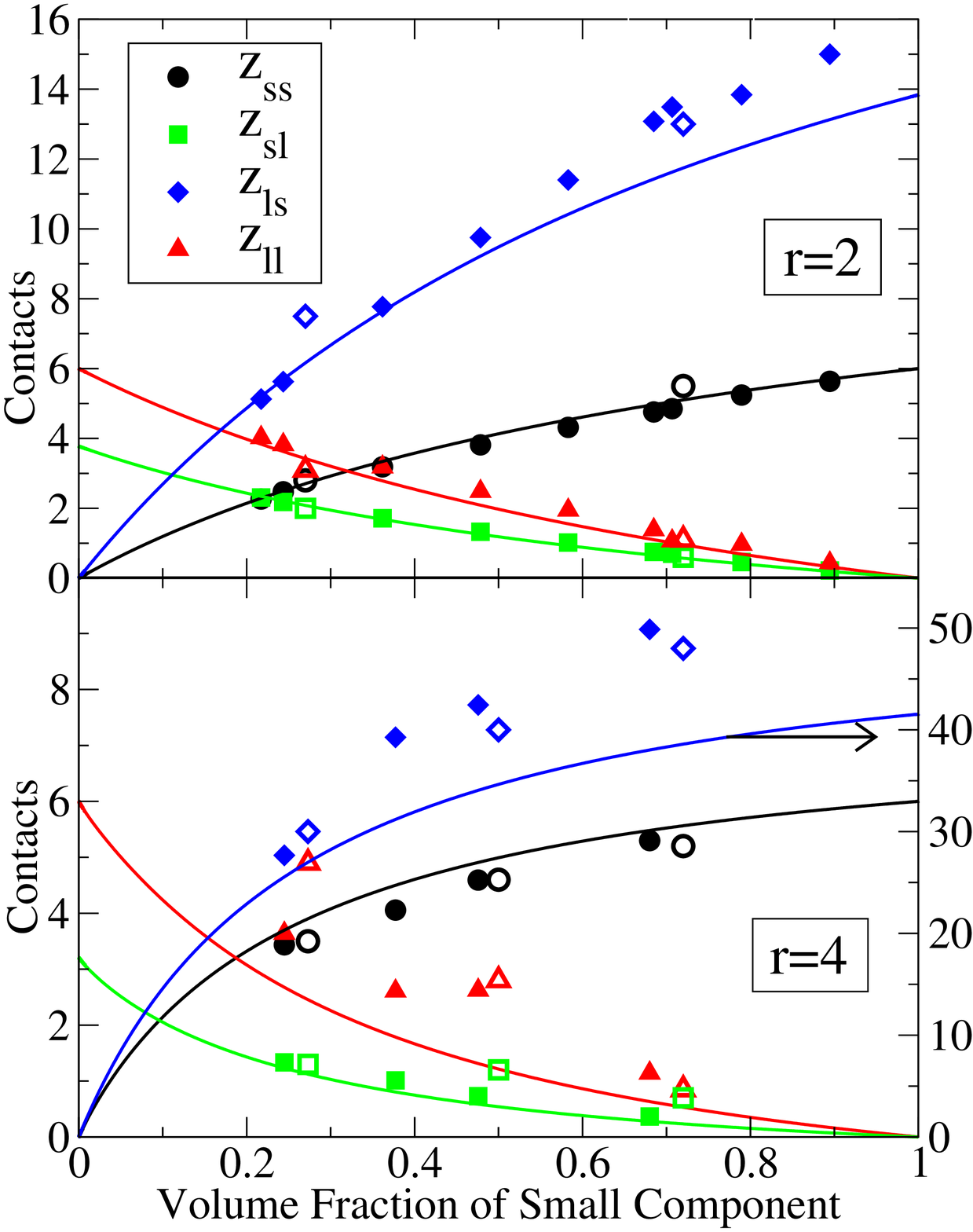}
\caption{
(Color online)
Partial average coordination numbers 
(small-small, small-large, large-small, large-large)
as a function of
volume fraction of the small particles $\h= 1/(1+x r^3)$ for different values of $r$.
Full symbols are numerical data from this work. Open symbols are experimental
data from Ref.~\cite{PZYZM98}.
Note that in the lower right panel a different scale is used for $z_{ls}$.
}
\label{fig:contacts}
\end{figure*}

{\it Results for binary mixtures} -
The details of the computation of the function $\SS(m,\f)$ for a general
multicomponent mixture, based on~\cite{PZ08}, can be found in \aldo.
Here we consider a binary mixture of two types of spheres $\mu=A,B$ in a volume $V$, 
with different diameter $D_\mu$ and density $\r_\mu = N_\mu/V$. 
We define $r = D_A / D_B > 1$ the diameter ratio and $x= N_A/N_B$ the concentration ratio;
$V_3(D) = \pi D^3/6$ the volume of a three dimensional
sphere of diameter $D$; 
$\f = \r_A V_3(D_A)+\r_B V_3(D_B)$ the packing fraction;
$\eta = \r_B V_3(D_B) / \f = 1/(1+ x r^3)$ the volume fraction of the small ($B$) component.

Once an equation of state for the liquid has been chosen,
the jamming packing fraction $\f_j$ is given in term of $\Si_j$ by
the solution of $\Si_j(\f)=\Si_j$.
The average coordination numbers at $\f_j$ are denoted $z_{\m\n}(\f_j)$, 
but we checked that their variations with $\f_j$ are negligible.
We used in $d=3$ the equation of state proposed in \cite{SLY05},
using the Carnahan-Starling equation for the monodisperse system~\cite{Hansen}.
The latter, as well as $\Si_j(\f)$ and $z_{\m\n}(\f_j)$, are given respectively in
Eqs.~(\ref{zetaadd}),
(\ref{SigmaLim}),
(\ref{eqcontatti}),
in \aldo.

{\it Numerical simulations} - We produced jammed packings of binary mixtures
of $N=1000$ 
hard spheres using the code developed by Donev et al.~\cite{DTS05c,DoWeb}.
In this algorithm spheres are compressed uniformly by increasing
their diameter at a rate $dD/dt = 2 \g$ while event-driven molecular dynamics
is performed at the same time.
In order to obtain a perfectly jammed final
packing, the later stages of compression must be performed very
slowly. On the other hand, at low density
slow compression is a waste of time, 
since the dynamics of the system is very fast.
Following~\cite{SDST06}, we find a good compromise 
by performing a four stages compression:
starting from random configurations at $\f=0.1$,
{\it i)} the first stage is a relatively fast
compression ($\g = 10^{-2}$) up to a reduced pressure $p = \b P/\r = 10^2$;
then we compress at {\it ii)} $\g = 10^{-3}$ up to $p = 10^3$;
{\it iii)}  $\g = 10^{-4}$ up to $p = 10^9$;
{\it iv)}  $\g = 10^{-5}$ up to $p = 10^{12}$.
The first stage terminates at a density $\f \sim 0.6$, and is fast enough 
to avoid crystallization and phase separation. During the following stages
the system is already dense enough to stay close to the amorphous structure
reached during the first stage. Still, little rearrangements (involving many
particles) are possible and allow to reach a collectively jammed final 
state~\cite{SDST06}. In the final configurations, we observe a huge gap between
contacting (typical gap $\sim 10^{-11} D$) and non-contacting (typical gap
$\gtrsim 10^{-6} D$) particles. We then say that two particles are in contact
whenever the gap is smaller than $10^{-8}D$.
Typically, a small fraction ($\lesssim 5 \%$ of
the total) of rattlers, \ie particles having less then 4 contacts,
is present. Once these are removed, the configuration is isostatic
(the total number of contacts is $6 N$)
within $1 \%$ accuracy.

{\it Comparison of theory and numerical/experimental data} -
During the four stages of compression, the pressure initially
follows the liquid equation of state up to some density close to
the glass transition density~\cite{PZ08,GV03,FGSTV03}.
Above this density, pressure increases faster and diverges
on approaching jamming at $\f_j$. The exact point where this happens
depends on compression rate.
This is a nice confirmation of a prediction of the theory, 
that different glassy states jam at different density; it was already
observed in~\cite{DTS06,SDST06} and recently discussed in great
detail in~\cite{BW09,HD09}.
Within the theory $\f_j$ is related to $\Si_j$, the value of configurational entropy at which
the system falls out of equilibrium; hence there is one free parameter, $\Si_j$, that depends 
on the compression protocol.
The equation of state of the glass obtained numerically with our protocol 
corresponds within our theory to $\Si_j \sim 1.5$. We decided to
use the value $\Sigma_j = 1.7$ that gives the best fit to the numerical data.
This is consistent with previous observations, that the configurational entropy
is close to $1$ when the system falls out of equilibrium~\cite{AF07}.
A detailed discussion of the behavior of pressure can be found in \aldo.

In figure~\ref{fig:phij}, we report the jamming density for different mixtures, putting
together our numerical results and experimental data from Ref.~\cite{YCW65}, and the theoretical results.
Note that a single ``fitting'' parameter $\Si_j$, that is strongly constrained, allows
to describe different sets of independent numerical and experimental data.
The prediction of our theory are qualitatively similar to previous ones~\cite{Do80,OT81},
but the quantitative agreement is much better. Interestingly, a similar qualitative behavior
for the glass transition density has been predicted in~\cite{GV03,FGSTV03}; although there is no {\it a priori}
reason why the jamming and glass transition density should be related~\cite{BW09}, it is reasonable to
expect that they show similar trends~\cite{FGSTV03}.

Finally, in figure~\ref{fig:contacts} we report the average
partial contact numbers for different mixtures. These values have been obtained
by removing the rattlers from the packing. As discussed above, the total coordination
is close to the isostatic value $z=6$. This is also a non trivial
prediction of the theory, see \aldo. As it can be seen from figure~\ref{fig:contacts},
the computed values agree very well with the outcome of the numerical simulation, at
least for $r$ not too large. Some discrepancies are observed in the contacts of the large
particles for large $r$. We produced packings of $N=10^4$ and checked that these are
not finite size effects. Also, inspection of the configurations seem to exclude the presence
of phase separation. However, for these values of $r$ and $x$, a large fraction of 
rattlers ($\sim 10\%$) is present within the small particles. This might affect the determination
of the partial contacts. It would be interesting to check if better results are obtained
using different algorithms. Experimental data from~\cite{PZYZM98} are also reported
in the right panel of figure~\ref{fig:contacts}.

{\it Conclusions} -
In this \piero we have extended our theory of amorphous packings to binary mixtures and have
tested it against numerical and experimental data. In particular
we have shown that the theory correctly predicts the variation of total density (or porosity)
and local coordination with mixture composition. We have also shown that the behavior of pressure
during compression follows the predictions of the theory. A striking prediction of the theory
is that different compression procedures lead to different final densities, which seems to be
confirmed by numerical data, see also \cite{BW09,HD09}. Note that the only free parameter is the value
of $\Si_j$, that is still strongly constrained (it must be close to 1). It affects slightly the values
of density (by varying $\Si_j$ in the reasonable range one can change $\f_j$ of $\sim 10\%$, see
figure~\ref{fig:pressure} in \aldo) and does not affect at all the curves in figure~\ref{fig:contacts} for
the local coordinations. As stated in the introduction, we believe that these results constitute a stringent
test of the idea that amorphous packings can be considered as the infinite pressure limit of metastable
glassy states. Note that our results have no implications on the existence of an ideal glass transition~\cite{SK00,DTS06}.
Indeed, we only used numerical data obtained using fast compressions. It is possible that using much slower
compressions physics changes dramatically and the glass transition is avoided. Still, these time scales are
out of reach of current algorithms and experimental protocols.

{\it Acknowledgments} -
We wish to thank A.~Donev for his invaluable help in using his code, and L.~Berthier for many important
suggestions and for providing equilibrated configurations at high density.

\bibliography{HS}

\bibliographystyle{mioaps}

\begin{widetext}

\newpage

\section*{\large \aldo}

\section{Behavior of the pressure during compression}

\begin{figure}[h]
\includegraphics[width=8cm]{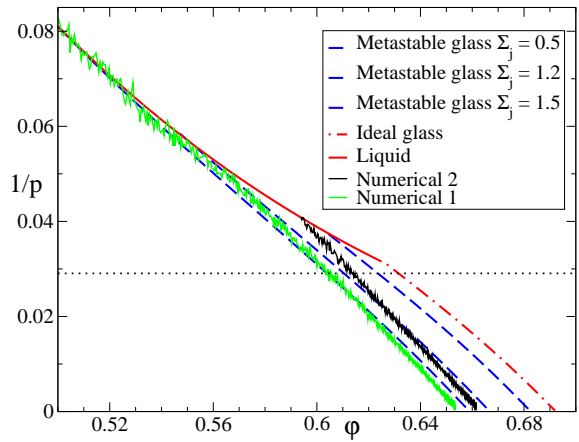}
\caption{
(Color online)
Inverse reduced pressure, $\rho/(\b P)$ as a function of the packing fraction $\f$ for
a mixture with $r=1.4$ and $x=1$.
Numerical data are obtained using two different protocols. In the first, compression
is started at low density. In the second, compression is started from an equilibrated configuration 
at $\f=0.58$.
The equation of state of different metastable glasses, corresponding to different $\Si_j$, are
reported as dashed lines. The dot-dashed line is the pressure of the ideal glass, corresponding
to $\Si_j = 0$. A numerical estimate~\cite{BW09} of the Kauzmann pressure, $p_K = 34.4$, is reported
as a dotted horizontal line.
}
\label{fig:pressure}
\end{figure}

Here we discuss in detail the behavior of pressure during compression in our numerical simulations.
In figure~\ref{fig:pressure} we report the evolution of the inverse
reduced pressure
during the first and second stages of compression ({\it Numerical 1}) for a mixture
with $x=1$ and $r=1.4$.
We observe, as in similar studies~\cite{DTS06,SDST06,BW09,HD09}, that the
pressure follows the liquid equation of state up to some density, that for this
system is around $\f\sim 0.56$. At this value of density, the relaxation time
of the liquid becomes long enough that liquid relaxation is effectively frozen 
on the compression time scale used in this work and the
system falls out of equilibrium. 
Although we did not measure relaxation time directly, this has been done in related
works that confirmed the correctness of this statement~\cite{BW09,HD09,FGSTV03}.
Above this density, pressure increases faster and diverges
on approaching jamming around $\f_j \sim 0.65$. The numerical equation of state
is compared with that of a glass state corresponding to $\Si_j = 1.5$.
This is consistent with previous observations, that the configurational entropy
is close to $1$ when the system falls out of equilibrium~\cite{AF07}.
In the same plot, we report the curve ({\it Numerical 2}) 
obtained starting the first stage 
from a carefully equilibrated
liquid configuration of the same mixture at $\f = 0.58$, 
kindly provided by L.~Berthier
(see \cite{BW09} for details on how this configuration was produced and equilibration was checked). 
In this case, since the relaxation time of the liquid at that density is already very long compared to our compression rate,
the system falls immediately out of equilibrium and
the pressure increases fast until jamming occurs at a higher density compared to the previous case.
In our interpretation, this corresponds to a glassy state with lower $\Si_j$
(compare with the theoretical curve for $\Si_j \sim 1.2$). This is a nice confirmation
of a prediction of the theory, that different glassy states jam at different density.
Finally, we report, for the same system, 
the numerically extrapolated value of the ideal glass transition pressure,
$p_K = 34.4$, see Ref.~\cite{BW09} for details of the careful extrapolation procedure. 
Again, this corresponds well (within 10$\%$)
to the computed value $p_K = 31.8$ from the
theory, see figure~\ref{fig:pressure}. Note that this coincidence 
does not prove the existence of the Kauzmann
transition, since on much larger time and length scales than the ones explored in these simulations,
a crossover to a non mean-field behavior might happen. Still, the coincidence shows that, 
{\it on the length and time scales explored by current numerical simulations}, the full mean field
phenomenology is observed, including the apparent extrapolation of Kauzmann transition.
Whether this transition really exists remains a major open point of the field, that however does not
affect the results presented here.

\section{Replica theory for multicomponent mixtures}

Here we show in detail how to generalize the computation
of \cite{PZ08} to multicomponent mixtures. We will not repeat
the discussion of \cite{PZ08} but we only explain how to modify
it for mixtures. Reading Appendix B and C and section VII of
\cite{PZ08} is necessary to follow the discussion.

We consider a multicomponent system in a (large) volume $V$ 
with partial densities $\r_\m = N_\m/N$. 
The total density is $\rho = \sum_\m \r_\m$ and we define $x_\m= \r_\m/\r$. 
Spheres of type $\m$ have
diameter $D_\m$ and we define $D_{\m\n}= (D_\m + D_\n)/2$. 
We denote by $\Omega_d = 2\pi^{d/2}/\G(d/2)$ the $d$-dimensional solid angle,
and by $V_d(D) = \pi^{d/2} D^d / \G(1+d/2)$ the volume of a $d$-dimensional
sphere of diameter $D$.
The hard
sphere potential $\phi_{\m\n}(r)$ is infinite for $r < D_{\m\n}$ and
zero otherwise: we denote $\chi_{\m\n}(r) = \exp[-\phi_{\m\n}(r)]$.
$g_{\m\n}(r)$,
as usual, is the $\m$-$\n$ pair correlation function~\cite{Hansen}. We will
use the shorthand notations $g_{\m\n} \equiv g_{\m\n}(D_{\m\n})$ for the 
contact values of $g_{\m\n}(r)$, and $V_d^{\m\n} = V_d(D_{\m\n})$ for the volume
of a sphere of diameter $D_{\m\n}$.

As discussed in \cite{CMPV99},
we assume that in the replicated liquid molecules are built
of particles of the same type. This amounts to assume that in
a glassy state particles of different type cannot easily exchange.
However, as the diffusion constant is always finite in a glass,
this assumption is wrong, since particles can exchange also within
a state. This is all the more true in the liquid
phase before the glass transition. 
Therefore, in order to correct
this error, we will subtract the mixing entropy from the configurational
entropy we will compute in the following, in order to obtain the correct
physical result.

According to \cite{CMPV99}, 
to each molecule of the replicated liquid we can attach a label
$\mu=1,\cdots,n$ according to the type of particles that build the molecule.
We denote coordinates in a molecule by $\bar x=(x_1,\cdots,x_m)$.
We have then to compute the free energy functional of a liquid made
of $n$ species of molecules, with a single molecule density 
$\r_\m(\bar x)$ and an interaction 
$\bar \chi_{\m\n}(\bar x,\bar y) = \prod_{a=1}^m \chi_{\m\n}(x_a-y_a)$.

\subsection*{Effective liquid}

We assume that the vibrations of the $m$ copies 
are described by a Gaussian distribution, corresponding to harmonic
vibrations:
\beq
\label{rrho}
\r_\mu(\bar x) = \r_\m \int d X \prod_{a=1}^m 
\frac{e^{-\frac{(x_a-X)^2}{2A_\mu}}}{(\sqrt{2\p A_\mu})^d} \ .
\eeq
The widths $A_\mu$ are variational parameters and we will maximize the entropy
with respect to them at the end.
We choose a replica (say replica 1) as a reference and consider the
vibrations of the other $m-1$ particles around the reference one.
We want to construct an expansion assuming that $A_\mu$ is small.
For $A_\mu = 0$, the $m-1$ copies coincide 
with the reference one, and $\SS(m,\f)$ is given by
the entropy $S(\f)$ of the non-replicated liquid 
plus the entropy of $m-1$ harmonic oscillators of spring constant 
$A_\mu$ (averaged over the concentration of the different species):
\beq
\SS^{(0)}(m,\f) = S(\f) + \sum_\m x_\m S_{harm}(m,A_\mu) \ .
\eeq
The harmonic part of the entropy is computed straightforwardly from
the term $\sum_\m \int d\bar x \r_\m(\bar x) [1 - \log \r_\m(\bar x)]$
in the free energy functional, see section V of \cite{PZ08}, where the
function $S_{harm}(m,A)$ is also defined.

A first order approximation is obtained by considering the effective
two-body interaction induced on the particles of replica 1 by the coupling
to the $m-1$ copies. 
This can be justified on the basis of a diagrammatic expansion
following the derivation in Appendix B of \cite{PZ08}, see also \cite{SW84}.
It is possible to show~\cite{Hansen}
that the diagrammatic expansion of the free energy functional for a 
multicomponent liquid is the same as the one of a simple liquid, provided
a label $\mu_i$ is attached to each vertex $i$, and in such a labeled diagram
$\r_{\m_i}(x_i)$ is placed on each vertex, $\chi_{\m_i\m_j}(x_i-x_j)$ on each link, and
a sum over $\mu_i$ is performed in addition to the integration over $x_i$. The same is
true for the molecular liquid. Hence, the treatment of diagrams performed in Appendix
B of \cite{PZ08} can be repeated exactly for a mixture, taking into account the presence
of the additional labels. This is straightforward and leads to the introduction of effective
potentials that now depend on the type of particles involved. In particular,
one obtains:
\beq\begin{split}
e^{-\phi^{eff}_{\m\n}(x_1-y_1)} &= \int dx_2 \cdots dx_m dy_2 \cdots dy_m
\r_\m(\bar x) \, \r_\n(\bar y)
 \prod_{a=1}^m e^{-\phi_{\m\n}(x_a - y_a)} \\
&\equiv e^{-\phi_{\m\n}(x_1 - y_1)} [1+ Q_{\m\n}(x_1-y_1)]  \ .
\end{split}\eeq

A first order approximation to $\SS(m,\f)$ is then obtained by substituting to the
entropy $S(\f)$ of the simple hard sphere mixture the free energy
of a liquid of particles
interacting via the potential $\phi^{eff}_{\m\n}(r)$:
\beq\label{S1}
\SS^{(1)}(m,\f) = -F[\f,\phi^{eff}_{\m\n}(r)] + \sum_\m x_\m S_{harm}(m,A_\m) \ .
\eeq
It is now evident that we can obtain better approximations of the true function
$\SS(m,\f)$ by considering also the three body interactions induced on particles of
the replica 1, and so on. Here, for simplicity, we will limit ourselves to consider the 
two-body interactions. 
To the same approximation, the pair correlation function $g_{\m\n}(r)$ of the system 
is given by the correlation of the effective liquid 
of particles interacting via $\phi_{\m\n}(r)$.
For small $A_\m$, $\phi^{eff}_{\m\n}$ is close to $\phi_{\m\n}$ and the free energy in Eq.~(\ref{S1})
can be computed in perturbation theory around the normal liquid~\cite{Hansen}, which
is taken as an input for the calculation of the properties of the glass.

\subsection*{Small cage expansion}

Now we follow the discussion in section VII of \cite{PZ08} to compute
the entropy of the effective liquid.
We assume that the cage radius is small, therefore $Q_{\m\n}$ can be
considered as a perturbation, and one has 
$\phi^{eff}_{\m\n}(r) \sim \phi_{\m\n}(r) - Q_{\m\n}(r)$.
Using the general relation~\cite{Hansen}
\beq
\frac{\d F[\f,\phi_{\mu \nu}(r)]}{\d \phi_{\mu \nu}(r)} = \frac{\r_{\mu}\r_{\nu}}{2\r} g_{\mu \nu}(r) \ ,
\eeq 
we obtain
\beq
-F[\f,\phi^{eff}_{\m\n}(r)] \sim S(\f) + 
\sum_{\m,\n} \frac{\r_{\mu}\r_{\nu}}{2\r} \int dr g_{\m\n}(r) Q_{\m\n}(r) \ ,
\eeq
where $S(\f)=-F[\f,\phi_{\m\n}(r)]$ is the entropy of the liquid.
The function $Q_{\m\n}(r)$ is different from zero only for $r -D_{\m\n} \sim O(\sqrt{A_\m+A_\n})$.
If $\sqrt{A_{\m}+A_{\n}} \ll D_{\m\n}$ we can assume that the function $g_{\m\n}(r)$
is essentially constant on the scale $\sqrt{A_{\m}+A_{\n}}$, then 
$g_{\m\n}(r) \sim g_{\m\n} \chi_{\m\n}(r)$ 
for $r -D_{\m\n} \sim O(\sqrt{A_\m+A_\n})$, and
\beq\begin{split}
\int dr g_{\m\n}(r) Q_{\m\n}(r) \sim g_{\m\n} 
\int dr \chi_{\m\n}(r) Q_{\m\n}(r) \ .
\end{split}\eeq

To compute the previous expression, all the considerations of Appendix C of \cite{PZ08}
can be repeated. Instead of a single function $Q(r)$, one has $Q_{\m\n}(r)$, and in the
two Gaussians one has $A_\m$ and $A_\n$ instead of $A$. 
At the very beginning of the discussion, then, one finds that it is enough to replace
$2 A \to A_\m + A_\n$ and follow the same steps.
We obtain
\beq
\int dr \chi_{\m\n}(r) Q_{\m\n}(r) = \frac{d \, V_d^{\m\n}}{D_{\m\n}}
\sqrt{2( A_\m+A_\n)} \, Q_0(m) \ .
\eeq
where $Q_0(m)$ is defined in Appendix C of \cite{PZ08}.

Putting all together, we find the final result
\beq\begin{split}
\SS(m,\f;\{A_\m\}) =  \sum_\m x_\m S_{harm}(m,A_\m) + S(\f) + 
\sum_{\m,\n} \frac{\r_{\mu}\r_{\nu}}{2\r}  g_{\m\n} \frac{d \, V_d^{\m\n}}{D_{\m\n}}
\sqrt{2( A_\m+A_\n)} \, Q_0(m)
\end{split}\eeq
This expression generalizes the result of section VII of \cite{PZ08}
to multicomponent mixtures; it has to be optimized over all the $A_\m$
to get the replicated entropy $\SS(m,\f)$.

\subsection*{Replicated free energy for binary mixtures}

The optimization with respect to $A_\m$ leads to $n$ coupled equations
that are not easy to solve in general. Therefore in the following we
focus on the case of binary mixtures that is of interest here.

For convenience we denote $A_A = A$ and $A_B = B$ the cage radii of the
two species.
We then obtain
\beq
\begin{split}
\frac{\partial\SS(m,\f;A,B)}{\partial A}&= \frac{\r_A^2 g_{AA} V_d^{AA} d}{2 \r D_{AA}}\frac{Q_0(m)}{\sqrt{A}}
+\frac{\r_A\r_B  g_{AB} V_d^{AB} d}{\r D_{AB}}\frac{Q_0(m)}{\sqrt{2(A+B)}}-
 x_A \frac{d(1-m)}{2 A}=0\\
\frac{\partial\SS(m,\f;A,B)}{\partial B}&= \frac{\r_B^2 g_{BB} V_d^{BB} d}{2 \r D_{BB}}\frac{Q_0(m)}{\sqrt{B}}
+\frac{\r_A\r_B  g_{AB} V_d^{AB} d}{\r D_{AB}}\frac{Q_0(m)}{\sqrt{2(A+B)}}-
 x_B \frac{d(1-m)}{2 B}=0\\
\end{split}
\eeq
Defining $\d = B/A$, $x= \r_A/\r_B$ and $r=D_A/D_B$,
we obtain the following equation for~$\d$:
\beq
\label{eqdelta}\begin{split}
&x \, r^{d-1} \, g_{AA} - \sqrt{\delta} \, g_{BB} 
+ \frac{\sqrt{2} (1-x \d)}{\sqrt{(1+\delta)}} \left(\frac{1+r}2 \right)^{d-1} g_{AB} =0 \ ,
\end{split}\eeq
and from its solution we obtain the optimal values of $A$ and $B = \d \, A$:
\beq\label{Astar}
\begin{split}
&\sqrt{A^{\star}(\delta)}
= \frac{1-m}{2 Q_0(m)}\frac{\la D^d \ra (1+x)}{2^{d} \, \f \, \G} \\
&\Gamma=\frac{1}{2} x g_{AA}D_{AA}^{d-1}+g_{AB}\frac{D_{AB}^{d-1}}{\sqrt{2(1+\delta)}}
\end{split}\eeq
where the packing fraction is $\f = \r_A V_d(D_A)+\r_B V_d(D_B)$
and $\la D^p \ra = \sum_\mu x_\mu D_\mu^p$.
Substituting these in $\SS(m,\f;A,B)$, we finally get
\beq\begin{split}
&\SS(m,\varphi)=S(\varphi)-\frac{d}{2}(1-m-\log m)-
 \frac{d}{2}(1-m)\log(2\pi A^{\star})-
  \frac{d}{2} \frac{1}{1+x} (1-m)\log (\delta)+
  \frac{2^d d \, \varphi}{\la D^d\ra}\frac{Q_0(m) \sqrt{A^{\star}} \Delta}{(1+x)^2} \\
&\Delta=x^2 g_{AA}D_{AA}^{d-1}+x g_{AB}D_{AB}^{d-1}\sqrt{2(1+\delta)}+g_{BB}D_{BB}^{d-1}\sqrt{\delta}
\end{split}\eeq
From this expression one can compute the complexity
using Eq.~(\ref{legm})
\cite{PZ08};
here it is not useful to report the complete expression. However, it is interesting to report
the following explicit expressions:
\beq\label{SigmaLim}
\begin{split}
\Sigma_j(\f) &= \lim_{m\to 0} \Sigma(m,\f) =
S(\varphi)-d \log\left(\frac{\sqrt{2}\left<D^d\right>(1+x)}{2^d\varphi \Gamma}\right)
-\frac{d}2 \frac{1}{1+x}\log \delta +\frac{d}{2} \ , \\
\Sigma_{eq}(\f) &= \Sigma(1,\varphi)=S(\varphi)-d\log \left[\sqrt{\frac{\pi}{2}}\frac{\left<D^d\right>(1+x)}{2^d K\varphi\Gamma}\right]-\frac{d}{2}\frac{1}{1+x}\log \delta
\end{split}\eeq
with $K=-Q'(1)=0.638\ldots$, see \cite{PZ08} for details.
From the latter expressions one can compute the Kauzmann density $\f_K$ which
is the solution of $\Si_{eq}(\f)=0$, and the Glass Close Packing density
$\f_{GCP}$ which is the solution of $\Si_j(\f)=0$~\cite{PZ08}.
More generally, given a set of metastable glasses of complexity $\Si_j$, under the
so-called {\it isocomplexity} assumption~\cite{PZ08}, one can compute their
jamming density as the solution of $\Si_j(\f)= \Si_j$. The results of this computation
for binary mixtures are discussed in the main text.
One should keep in mind that, as discussed above, Eqs.~(\ref{SigmaLim}) incorrectly includes
the mixing entropy (coming from the $S(\f)$ term). This has to be subtracted in order
to get the correct physical result.

\subsection*{Coordination numbers}

Here we show how to compute the partial coordination numbers. The technical part of the
computation follow closely the derivation in \cite{PZ08}, hence the only nontrivial point
is to add indices corresponding to particle types.
As discussed in section VII.C.3 of \cite{PZ08}, the integral of the glass correlation function,
$\wt g^{\mu\nu}(r)$, on a shell $D_{\mu\nu} \leq r \leq D_{\mu\nu} + O(\sqrt{\f_j-\f})$ gives
the number of particles of type $\nu$ 
that are in contact with a given particle of type $\mu$ for $\f \to \f_j$.
A straightforward generalization of the derivation of section VII.C.2
of \cite{PZ08} shows that, at the leading order close to contact,
$\wt g_{\m\n}(r) = g_{\m\n}(r) [1 +Q_{\m\n}(r)] \sim g_{\m\n}  [1 +Q_{\m\n}(r)]$.
Then one obtains
\beq\label{zdef}\begin{split}
z_{\mu \nu} = 
\Omega_d \rho_{\nu} D_{\mu\nu}^{d-1} g_{\mu\nu} \int_{D_{\mu\nu}}^{D_{\mu\nu}+O(\sqrt{(A_{\mu}+A_{\nu})/2})} dr \, [1+ Q_{\mu\nu}(r)] \ .
\end{split}\eeq
In the limit $\f \to \f_j$, one has $A_\m \propto m \to 0$ and the integral of $Q_{\m\n}(r)$ can be
easily evaluated~\cite{PZ08}. The result is
\beq\label{zdef2}\begin{split}
z_{\m\n}
=\Omega_d \rho_{\nu} D_{\mu\nu}^{d-1} g_{\mu\nu} \sqrt{2(a_{\mu}+a_{\nu})}
\end{split}\eeq
where $a_{\mu} = \lim_{m\to 0} A_{\mu} \left(\frac{Q_0(m)}{1-m}\right)^2$. 

In the explicit case of binary mixtures, note that the equation (\ref{eqdelta})
for $\d$ does not depend on $m$. Using also Eq.~(\ref{Astar}), we obtain
\beq\label{eqcontatti}
\begin{cases}
&z_{AA}=x D_{AA}^{d-1} g_{AA}\frac{d}{\Gamma} \\
&z_{BB}=D_{BB}^{d-1} g_{BB}\sqrt{\delta}\frac{d}{\Gamma} \\
&z_{AB}=D_{AB}^{d-1} g_{AB}\sqrt{\frac{1+\delta}2}\frac{d}{\Gamma} \\
&z_{BA}=xD_{AB}^{d-1} g_{AB}\sqrt{\frac{1+\delta}2}\frac{d}{\Gamma}
\end{cases}
\eeq
Note that it is possible to show, using the definitions of $\d$ and $\G$,
that the average total coordination $z = \sum_{\m\n} x_\m z_{\m\n} = 6$,
\ie the packings are predicted to be isostatic irrespective of the mixture 
composition.

\subsection*{Equation of state for liquid hard sphere mixtures}

We used a generalization of the Carnahan-Starling equation of state,
which is defined by the following relation for the contact value
of the radial distribution function~\cite{SLY05}:
\beq
\begin{split}
&g_{\mu\nu}(\varphi)=\frac{1}{1-\varphi}+\left(g_{pure}(\varphi)-\frac{1}{1-\varphi}\right)
\frac{\left<D^{d-1}\right>D_{\mu\mu}D_{\nu\nu}}{\left<D^d\right>D_{\mu\nu}} \ ,
\label{zetaadd}
\end{split}\eeq
where $g_{pure}(\varphi)$ is the contact value of $g(r)$ for the pure system,
given by the standard Carnahan-Starling equation~\cite{Hansen}.
The pressure is then given by the exact relation
\beq
p(\varphi)=\frac{\beta P}{\rho}=1+\frac{2^{d-1}}{\left<D^d\right>}\varphi\sum_{\mu\nu}x_{\mu}x_{\nu}D^d_{\mu\nu}g_{\mu\nu}=-\varphi\frac{\partial S(\varphi)}{\partial \varphi}
\eeq
Integrating this expression one obtains the entropy $S(\f)$. The additive integration constant
is fixed by the condition that, for $\f\to 0$, the entropy tends to the ideal gas value
$S(\f) = 1 - \log \r -  \sum_\m x_\m \log x_\m$. This expression includes the mixing entropy
that must be subtracted before substituting $S(\f)$ into Eqs.~(\ref{SigmaLim}).

\vskip30pt

\end{widetext}

\end{document}